\documentclass[aps,prper,twocolumn,superscriptaddress,letterpaper,longbibliography]{revtex4-1}

\usepackage{graphicx}
\usepackage{mdframed}
\usepackage{framed}
\usepackage{indentfirst}
\usepackage{natbib}
\usepackage{amsmath,amssymb}
\usepackage{subfigure}
\usepackage{enumitem}
\usepackage{multirow}
\usepackage{epstopdf}
\usepackage[T1]{fontenc}	
\usepackage[latin9]{inputenc}	
\usepackage{geometry}		
\usepackage[above,below]{placeins}	
\usepackage{times}

\usepackage{hyperref}  
\hypersetup{colorlinks=true,urlcolor=blue,citecolor=blue,linkcolor=blue}   
\usepackage{enumitem}          
\setlist{nosep}                 

\begin{document}

\begin{titlepage}

\title{Student behavior and test security in online conceptual assessment}



\author{Bethany R. Wilcox}
\affiliation{Department of Physics, University of Colorado, 390 UCB, Boulder, CO 80309}

\author{Steven J. Pollock}
\affiliation{Department of Physics, University of Colorado, 390 UCB, Boulder, CO 80309}

\begin{abstract}
Historically, the implementation of research-based assessments (RBAs) has been a driver of education change within physics and helped motivate adoption of interactive engagement pedagogies.  Until recently, RBAs were given to students exclusively on paper and in-class; however, this approach has important drawbacks including decentralized data collection and the need to sacrifice class time.  Recently, some RBAs have been moved to online platforms to address these limitations.  Yet, online RBAs present new concerns such as student participation rates, test security, and students' use of outside resources.  Here, we report on a pilot study addressing these concerns.  We gave two upper-division RBAs to courses at five institutions; the RBAs were hosted online and featured embedded JavaScript code which collected information on students' behaviors (e.g., copying text, printing).  With these data, we examine the prevalence of these behaviors, and their correlation with students' scores, to determine if online and paper-based RBAs are comparable.  We find that browser loss of focus is the most common online behavior while copying and printing events were rarer.We found no statistically significant correlation between any of these online behaviors and students scores.  We also found that participation rates for our upper-division population went up when the RBA was given online.  These results indicates that, for our upper-division population, scores on online administrations of these RBAs were comparable to in-class versions.  \clearpage
\end{abstract}

\maketitle

\end{titlepage}

\section{\label{sec:intro}Introduction \& Background}

Research-based assessments (RBAs) have become a cornerstone of physics education research (PER) due in large part to their ability to provide a standardized measure of students' learning that can be compared across different learning environments or curricula \cite{madsen2017rbaRL}.  As such, these assessments are a critical step along the path towards making evidenced-based decisions with respect to teaching and student learning.  RBAs have historically been a strong driver in promoting the need for, and adoption of, educational reforms in undergraduate physics courses (e.g., \cite{beichner2007scaleup, hestenes1987modeling, crouch2001pi}).  

However, despite their value, there are a number of barriers to wide scale implementation of RBAs \cite{wilcox2016admin}.  For example, most of the existing RBAs require that an instructor sacrifices 1-2 full class periods to administering the RBA pre- and post-instruction.  For many instructors feeling pressure to cover as much content as possible over the course of a semester this sacrifice is difficult to justify.  In addition to the demand for class time, instructors must also sacrifice valuable time outside of class to analyze their students' performance.  Many instructors are not experts in assessment and struggle with analysis and interpretation their students' scores.  

Recently, physics education researchers have attempted to address both of these challenges by shifting RBAs to online platforms (e.g., \cite{wilcox2016admin, vandusen2015lasso, PhysPortwebsite}).  Hosting the RBAs online allows instructors to assign the RBA for student to complete outside of class.  Moreover, since the online platform allows for easy standardization and centralization of the data collection, many of these systems also include a mechanism to automate the analysis of students' response so that the instructor no longer needs to perform this analysis themselves.  While the online systems have a lot of potential for encouraging more wide-spread use of RBAs, these systems bring with them an number of other concerns, particularly around the potential for reduced participation rates, students' use of outside resources, and breaches of test security.  

Previously, the Learning Assistant Student Supported Outcomes (LASSO) study has addressed concerns about changes in scores and participation rates between online- and paper-based RBAs.  They found that with respect to participation rates, differences between the two formats vanish when best practices are used for the online implementations \cite{jariwala2017lasso}.  These practices include: multiple email and in-class reminders, and offering extra or regular course credit for participating.  They also found that, when participation rates were similar, students' overall performance was also statistically comparable \cite{nissen2017lasso}.  These findings suggest that participation rates and equivalency of students scores between online- and paper-based RBAs are not a concern as long as instructors follow best practices for encouraging participation in the online RBA.  

Less work, however, has been done to investigate the validity of concerns about students' use of outside resources or breaches in test security.  We are aware of only one study addressing these issues conducted in the context of an introductory astronomy course \cite{bonham2008online}.  In this study, Bonham \cite{bonham2008online} used JavaScripts and other applets to detect when students engaged in behaviors like printing browser pages, coping or highlighted text, and switching in to other browser windows while taking an online astronomy concept assessment.  They found no instances of students printing pages, and only 6 cases (out of 559) they deemed were probable incidence of students copying text.  Students switching browser windows was more common; however, Bonham argued these events appeared random and were not systematically associated with particular questions.  There were several important limitations to Bonham's study.  In browsers other than Microsoft Explorer, copy events and save events were detected through the proxies of highlighting text and page reloads respectively.  As Bonham noted, highlighting text as a proxy for copying results in many false positives, and there was no discussion of how these behaviors related to performance on the RBA.  

Here, we replicate and extend the study by Bonham in the context of upper-division physics courses.  We limit our attention to upper-division RBAs and student populations as a pilot study primarily because upper-division students are typically more mature and motivated than introductory students.  Thus, we argue these students are most likely to interact with the online RBAs as we would want them to.   In other words, upper-division students represent the population with whom online RBAs are least likely to be misused, and thus provide a proof of concept for whether online RBAs are viable for any undergraduate population.  In the following section (Sec.\ \ref{sec:context}) we discuss the context and methods used in this study.  We then present our findings with respect to students' online behaviors when taking the RBAs as well as how these behaviors correlate with their overall performance (Sec.\ \ref{sec:results}).  Finally, we end with a discussion of our conclusions and limitations of the study (Sec.\ \ref{sec:discussion}).

\section{\label{sec:context} Context \& Methods}

The two upper-division RBAs used in this study were the Quantum Mechanics Conceptual Assessment (QMCA) \cite{sadaghiani2015qmca} and the Colorado Upper-division Electrostatics Diagnostic (CUE) \cite{wilcox2015cue}.  Both the QMCA and CUE are multiple-choice or multiple-response assessments targeted content from the first semester in a two-semester sequence in junior-level Quantum Mechanics, and Electricity and Magnetism respectively.  Both assessments were administered online, using the survey platform Qualtrics, during the final week of the regular semester.  Student responses were collected from six distinct courses at five institutions.  All five institutions are four-year universities spanning a range of types including two doctoral-granting institutions classified as very high research, one masters-granting institution classified as Hispanic-serving, and two bachelors-granting institutions.  The authors taught two of the six courses, and the remaining instructors volunteered.  

In all cases, the instructors offered regular course credit to their students for simply completing the RBA (independent of performance).  In most cases, students received multiple in-class reminders to complete the assessment, and participation rates by course varied from 82\% to 100\%.  Since the goal of this study is not to compare courses, the remainder of the analysis will consider these students in aggregate.  Overall, the courses enrolled 217 students of which 207 responded to the RBAs for an overall participation rate of 94\%.  This participation rate is somewhat higher than what has been observed for either paper-based or online RBAs in previous studies at the introductory level \cite{jariwala2017lasso, bonham2008online}.  We also have historical participation rates available for the two of the  courses in the data set.  Historical participation rates for these courses were never higher than 80\%, suggesting that the participation rate for these courses actually increased significantly when the RBA was given online.  We do not have consistent access to data on the racial or gender distributions for these students.  

On the first page of the assessment, students were instructed to complete the RBA in one sitting without the use of outside resources such as notes, textbooks, or Google.  To capture students' online behaviors, we embedded JavaScript code into the online prompts to look for instances of students copying text, printing from their browser, and clicking into another browser window.  In all cases, these behaviors were time stamped to determine when each action occurred and how many times each student exhibited that behavior.  This JavaScript code could only detect activities that happen at the browser level; activities at the computer level (e.g., taking a screenshot or clicking into another program) were not recorded by the code.  While such data would be useful, modern browsers nearly all have security features to prevent cookies and scripts in browsers from collecting information on activities happening outside the current browser window.  

For browser print commands (e.g., ``control-p") and copy text commands (e.g., ``control-c"), the only data collected were when and how often these commands were issued.  Data on browser focus were somewhat more complex.  The code was designed to listen for a change in browser focus, and then record whether the RBA tab was visible 4 seconds after the browser focus event occurred.  Thus, if a student clicks into a new browser tab and stays in that tab for more than 4 seconds, the code would record a browser focus event and tag it ``hidden."  A ``hidden" browser focus event means that the student left the RBA without returning to it within 4 seconds.  If a student clicked into another browser tab and then clicked back into the RBA within 4 seconds (and remained there for more than 4 seconds), the code would record two browser focus events -- one for the click out and one for the click in -- and would tag both as ``visible."   A ``visible" browser focus event means that the student returned to the RBA for more than 4 seconds after having left it for any amount of time.

In addition to the data on students' online behaviors, we collected students' scores on the assessment, and total duration between students starting and submitting the assessment.  Below, we examine these data to determine how prevalent specific online behaviors were for this population of students.  We also examine correlations between these behaviors and students' performance on the assessments overall.

\section{\label{sec:results}Results}

\subsection{\label{sec:print}Print Events}
The primary concern associated with students printing or saving RBAs is that these students might publicly post the assessment and thus breach the security of the assessment by making it available to other students.  In the full data set of 207 responses, only 3 students (or 1.4\%) had recorded print events; of these three, only one actually submitted the assessment (the others only opened the assessment and paged through all the questions).  One possible deterrent for students printing the assessments is that the online RBAs used here were designed to have the same number of pages as their paper counterparts and the pages are displayed one at a time.  Thus, a student wanting to save the whole RBA would need to print (or screenshot) every individual page of the assessment, rather than simply being able to save in one go.  However, for all three print cases in our data set, the student had at least 14 disctinct print events suggesting that they may indeed have printed each separate page of the assessment despite the relative tedium of doing so.  

Print commands themselves do not necessarily indicate a student who is intending to breach the security of the assessment.  In fact, one of the instructors (SJP) reported interacting with a student during help hours in which the students pulled up screenshots of the assessment which he had taken to study from after the fact.  The student made no attempt to hide the screenshots and was upfront with his motivation for taking the screenshots as a study tool.  Moreover, even if a student did post the RBA prompts online, without corresponding solutions, which were never released to the students, it is not clear that access to the RBA prompts alone actually represents a significant threat to the assessment's security or validity.  

To test for any immediate security breaches of the assessments, we Googled the prompts for each question on both the QMCA and CUE several weeks after the assessments had closed.  In no cases was there any indication that the item prompts or their solutions had been uploaded in a way that ranked high in Google's listing.  However, as Google's algorithm can change based on search patterns, it is likely necessary to do this type of verification periodically to ensure no solutions have surfaced.  In several cases, Googling the item prompts pulled up publications on the test itself.  In some cases, these publications included supplemental material which contained the grading rubrics for the assessment in one form or another (open-ended or multiple-choice).  It is worth noting that in all cases, these rubrics were buried at the end of a long publication or thesis and not clearly marked, and it is not clear if a student who was unfamiliar with the specific publications (or the nature of academic publication more generally) would be able to locate the rubrics without considerable persistence.  However, this suggests that the greatest threat to the security of these RBAs in an online format may actually be our own publications combined with the fact that the premier PER publication venue is open access.  

\subsection{\label{sec:focus}Browser Focus Events}
Online RBAs introduce a potential for students to become disengaged from the assessment in a way that is less likely in paper-based administrations.  Loss of browser focus is one proxy for students disengaging from the RBA.  Focus events were the most common events in the data set with a total of 124 (60\% of 207) students with at least one browser focus event in which their RBA window became hidden for more than 4 seconds.  For these students, we examined trends in the number and duration of browser focus events by grouping them to isolate sustained changes in browser visibility.  In other words, if a students' survey page becomes hidden, how long before it becomes visible again, independent of whether there are additional browser hidden events in between (indicating that the student clicked back into the survey window, but did not remain there for more than 4 seconds)?  Here we will report median and max durations as the presence of significant outliers makes the average less meaningful.  The median number of sustained browser hidden periods per student was 2.  Fifty-two students (42\% of the 124 with sustained hidden periods) had only one period, and 14 students (11\%) had 10 or more browser hidden periods.  The median duration of each loss of focus period was 32 seconds, and the maximum duration was 23.7 hours.  Additionally, 386 (67\% of 575) of the sustained browser hidden periods were 1 min or less in duration, and only 54 of the periods (9\%) were longer than 5 min.  This suggests that the majority of students in the data set did click out of the assessment tab; however, two-thirds of the time they were away from the RBA for no more than 1 minute, a time-frame comparable to how long a student might ``space out" during class.  

We also examined whether the appearance or duration of loss of focus events correlated with students' scores on the assessment.  For all courses, students with loss of focus events scored lower on average than other student by between 2\% and 8\% for a given individual class, and 3\% for the students in aggregate.  These differences were not statistically significant for any individual class or overall, indicating that students with loss of focus events scored similarly to those without.  Additionally, we examined the Spearman correlation coefficient between the total time students spent with their browser hidden relative to their score on the assessment and find a small and negative ($r=-0.1$) but not statistically significant correlation ($p=0.2$).  We selected the Spearman correlation because it is less sensitive to the presence of outliers than the other coefficients.  This result suggests that there is no significant association between amount of time spent away from the RBA and performance.  

\subsection{\label{sec:copy}Copy Events}
We had two overall concerns associated with students copying the text of items on RBAs.  The first is the same concern associated with print events (i.e., that students may copy item text in order to save and later post them in a public forum).  The second concern is that students may copy text in order to search the internet in an attempt to ``look up" the correct answer.  Copy events were more common in the data set than print events, with a total of 95 copy events made by 18 studets (9\% of the 207).  The number of copy events per student varied significantly with a median of 2 copy events per student (in the 18).  Five of the 18 students had only one copy event while one student had a full 31 distinct copy events.  Interestingly, the student with 31 copy events did not actually complete the assessment and only questions 1-5 were displayed before the survey was closed.  One possible explanation for this behavior might be that this student was attempting to save the assessment in another program (e.g., Word) and eventually gave up after the first few items.  Thus, none of the students in our data set utilized copy commands in a manner that could have threatened the security of the RBAs (though nothing prevented them from doing so).  

The second concern about copy events, as related to attempts to Google answers to the questions, cannot be directly detected from the copy events alone.  However, if a student copies text with the intention of Googling that text, this behavior would most likely be characterized by a copy event followed immediately by a sustained browser hidden focus event.  Of the 95 distinct copy events, 53 (56\%) were followed within 5 seconds by a loss of focus event where their survey window became hidden (i.e., they stayed in the new window for more than 4 seconds).  Of the remaining 42 copy events that were not followed by a loss of focus event, 31 were by the single student discussed above.  The final 11 copy events were typically characterized by either the first of two quick consecutive copy events followed by a single loss of focus event, or single copy events not connected temporally with a loss of focus event.  This indicates that a majority of copy events were immediately followed by the student switching into a new browser window and remaining there for more than 4 seconds, consistent with the pattern we would expect if they were trying to Google the item prompts.  

Given this pattern, we also examined whether the students with copy events had any difference in performance from other students.  Only two courses had more than one student with a copy event per class, making a statistical comparison of scores possible only for those courses.  In both cases, students with copy events scored lower than the rest of the class by close to 10\%, though small-N in the ``copy" category did not allow for sufficient statistical power for this difference to be statistically significant in either case.  This may indicate that it is likely the lower performing students who attempt to copy and google text, but that the process does not appear to increase their scores beyond the rest of the students.

\subsection{\label{sec:time}Time to Completion}
We also examine the total amount of time to completion for each student to determine whether student's scores are related to how long it took them to complete the assessment.  Total time data are calculated by comparing the recorded time when the student first opened the survey link to when they made their final submission of the survey.  This does not remove periods when browser focus was lost, and can even include a period when the survey window was closed and later reopened.  As such, these duration do not necessarily reflect the amount of time a student actually worked on the assessment, merely the amount of time that passed between them opening and submitting the assessment.

For the vast majority of students (74\%, 154 of 200) the total time between start and submit fell within a time frame of 15-60 min, consistent with what would be required of a student taking the RBA in class.  We can improve somewhat on the raw time data by subtracting out the total time for each student during which their survey window was hidden, suggesting they may not have been working on the assessment.  To determine if total time is related to performance, we calculate Spearman correlations between score and time to completion.  For both total time and time minus loss of focus, the correlation with score on the assessment was small ($r=0.01$ and $r=0.1$ respectively) and statistically insignificant.  This indicates that, consistent with what has been found for paper-based RBAs, there is no significant relationship between the amount of time spent on the RBA and students' scores.

\section{\label{sec:discussion}Discussion \& Limitations}
We collected online responses to two upper-division research-based assessments from six upper-division quantum mechanics and electricity and magnetism courses.  This work is part of ongoing research to determine whether students' performance on RBAs shifts when these assessments are given online.  For two of the courses in the data set, we also have historical scores from students in these same classes with the same instructor where the RBA was given on paper and during class.  Comparisons of the online and in-class scores showed the online scores being 4-5\% lower.  This difference was not statistically significant (two-tailed t-test, $p>0.05$) and was largely driven by the presence of a larger lower tail of the distribution in the two online administrations.  This, combined with the higher participation rates in the online administration suggests that administering RBAs online to an upper-division population encourages more of the lower performing students to participate.  

In addition to student responses to the assessments themselves, we also collected data using embedded JavaScript code on students' online behaviors such as copying text, printing browser pages, and loosing browser focus by clicking into other browser tabs.  We found that only a small number of students (roughly 2\%) printed or copied item prompts in a manner that suggested they were attempting to save some or all of the item prompts.  However, we have anecdotal evidence that at least some of these students were saving the prompts solely for their own future studying.  Thus, while our data suggest that some students do engage in printing and copying behaviors, we argue that the threat to the security of the assessment that they pose is not significantly worse than that introduced by handing out a classroom set of paper RBAs where a student could simply keep the handout.  

More students (roughly 65\%) engaged in online behaviors resulting in loss of browser focus and indicating that the students may have disengaged from the RBA for a period of time.  However, roughly two-thirds of the periods where students lost browser focus lasted less than 1 minute, and less than 10\% of the periods lasted for longer than 5 minutes.  Moreover, the total amount of time that students spent away from their assessment tab did not correlate significantly with students' scores on the RBAs.  Thus, we argue that while the potential for distraction and disengagement certainly increases with online RBAs, our data suggest the majority of students do not become disengaged for long periods and that this disengagement does not appear to impact their performance.  

Evidence of copying text was observed in roughly 9\% of the students in our sample.  Roughly half of the copy events were immediately followed by a browser focus event in which the RBA tab became hidden.  Such a pattern is consistent with what we would observe if students were attempting to Google the item prompts in an attempt to determine the correct answers.  While it is not possible for us to determine for certain if that is what the students were doing, the pattern is suggestive.  However, comparisons of these students' scores with the students who did not copy text show that students with copy events had lower but statistically comparable performance on the RBA.  This suggests both that it may be the students who are struggling more who are more inclined to try to use outside resources, and that even if they are doing so, it does not appear to have a noticeable positive impact on their performance.  

The work presented here has some important limitations.  The code that captured students' online behaviors can only detect actions at the browser level, meaning that actions at the computer level (like switching into a new program) cannot be detected.  Moreover, we restricted our focus to upper-division students because we believe they are the population most likely to engage with the online RBAs in an authentic way.  For both of these reasons, our data should be interpreted as a lower bound on the appearance of these behaviors.  Extending this study to additional upper-division courses as well as to introductory courses will be the subject of future work.  However, these pilot results do suggest that, for upper-division courses, online assessment is a promising alternative that brings with it many potential logistical advantages.

\begin{acknowledgments}
This work was funded by the CU Physics Department.  Special thank you to the faculty and students participants and the members of PER@C for all their feedback.  
\end{acknowledgments}

\bibliography{master-refs-01-19}

\end{document}